\documentclass[pra,aps,twocolumn,tightlines,showpacs]{revtex4}

\usepackage[]{times,amsmath,epsfig,graphicx}

\begin{document}

\title{Hard and Soft Excitation Regimes of Kerr Frequency Combs}

\author{A. B. Matsko, A. A. Savchenkov, V. S. Ilchenko, D. Seidel, and L. Maleki}

\affiliation{OEwaves Inc., 465 N. Halstead St. Ste. 140, Pasadena, CA 91107}

\begin{abstract}
We theoretically study the stability conditions and excitation regimes of hyper-parametric oscillation and Kerr frequency comb generation in continuously pumped nonlinear optical resonators possessing anomalous group velocity dispersion. We show that both hard and soft excitation regimes are possible in the resonators. Selection between the regimes is achieved via change in the parameters of the pumping light.
\end{abstract}

\pacs{42.62.Eh, 42.65.Hw, 42.65.Ky, 42.65.Sf}

\maketitle

The phenomenon of four-wave mixing (FWM) results in hyper-parametric oscillation  \cite{kippenberg04prl,savchenkov04prl} and frequency comb generation \cite{delhaye07n,savchenkov08prl,kippenberg11s} in solid state optical microresonators. The oscillation is closely related to  modulation instability lasing  \cite{nakazawa89jqe1,haelterman92ol,coen97prl,coen01ol} and
parametric oscillation in optical fibers \cite{serkland99ol,sharping02ol,matos04ol,deng05ol,xu09leos}.
Four-wave mixing results in  generation of frequency sidebands in the light leaving the resonator. The sidebands are separated approximately by integer numbers of free spectral ranges (FSRs) of the resonator and are always equidistant ($\tilde \omega_+-\omega=\omega-\tilde \omega_-$). This is because the oscillation results from the four photon process $\hbar \omega + \hbar \omega \rightarrow \hbar \tilde \omega_+ + \hbar \tilde \omega_-$, where $\omega$, $\tilde \omega_+$ and $\tilde \omega_-$ are the frequencies of the pump light and the generated sidebands, respectively. As the power of the pump light is increased  multiple sidebands are generated and  the Kerr frequency comb is produced \cite{delhaye07n}.

The efficiency of the FWM process depends on group velocity dispersion (GVD) of the resonator. A resonator with nonzero GVD does not have an equidistant spectrum, since $2\omega_0-  \omega_{+}-  \omega_{-} = c\beta_2\omega_{FSR}^2/n_0$, where $\beta_2$ is the GVD (by definition, normal dispersion corresponds to $\beta_2>0$),  $n_0$ is the refractive index of the material, assumed to be constant in the calculations; $c$ is the speed of light in the vacuum, $2 \omega_{FSR} \approx \omega_+-\omega_-$ is the FSR of the resonator; $\omega_0$, $\omega_+$, and $\omega_-$ are the eigenfrequencies of the consecutive modes.

In addition to  dispersion, the spectrum of the resonator is also influenced by self and cross phase modulation effects resulting from the nonlinearity of the material. In the case of anomalous GVD and positive cubic nonlinearity, these effects compensate each other so the spectrum of the resonator becomes locally equidistant for a particular power and wavelength of light within the resonator. The inequidistance of the modes cannot be compensated in the case of normal GVD. The frequency difference between the modes of the pumped resonator influences the FWM and impacts the spectrum and threshold of the hyper-parametric oscillation and frequency comb generation.

The Kerr frequency combs excited in optical microresonators are promising for many practical applications since their spectral width can span an octave \cite{delhaye11prl,okawachi11ol}, and their frequency stability is extremely high \cite{kippenberg11s}. Understanding the fundamental properties of these combs is necessary to achieve their optimal performance. This is why, in addition to  multiple experimental investigations, the resonant FWM and Kerr comb generation have been studied theoretically. For example, it was shown that additive modulational instability ring lasers can operate in both normal and anomalous GVD regimes \cite{haelterman92ol}. This result was adjusted to describe the nonlinear processes in microresonators  \cite{matsko05pra,matsko09sfsm,matsko11nlo}. Theories supporting the idea of preferred generation of the optical Kerr comb in resonators possessing anomalous GVD were developed \cite{agha07pra,agha09oe}. An analytical expression describing the optical pulses generated in resonators with anomalous GVD was derived \cite{matsko11ol}. Finally, generation of Kerr frequency comb was investigated via numerical simulations \cite{chembo10prl,chembo10pra}.

The aim of this paper is to analyze the dependence of the excitation dynamics of the hyper-parametric oscillation and Kerr frequency combs depending on the power and frequency of the external continuous wave pump. The dynamics of Kerr comb formation was studied previously \cite{chembo10pra} with approximation of small amplitude of the comb frequency harmonics. This assumption allowed modeling the soft regime of the comb excitation. In this work we solved the problem without such an approximation, analyze the oscillation onset in the microresonators, and s that both hard and soft excitation of the oscillation is possible.

The  soft oscillation onset is reached when pump photons are not initially present in the resonator. Here, the growth of the oscillation sidebands occurs adiabatically. The hard onset of the oscillation occurs with a discontinuous jump of the intensity of  oscillation sidebands to a certain finite level, at the threshold. The hard excitation occurs only when  pump photons are initially present in the resonator. The steady state solution corresponding to the hard excitation cannot be reached adiabatically. In other words,  slow modifications of any parameter of  the resonator or the pump cannot bring the system to an stable solution requiring hard excitation.  Here slow means a time shorter than  the lifetime of the light confined in the resonator.

We solve the set of nonlinear differential equations describing the hyper-parametric oscillation \cite{matsko05pra} and frequency comb generation \cite{chembo10pra} in steady state, and analyze the stability of the solutions in a microresonator possessing a net anomalous GVD. While the first order hyper-parametric oscillation (only the continuous wave optical pump and the first pair of sidebands are involved in the process) is described analytically, the higher order oscillations are simulated numerically.

We found that oscillation sidebands can be much smaller than the pump in the case of soft excitation, while they are comparable with the pump in the case of hard excitation. The sidebands generated in the case of hard excitation produce  optical pulses that travel in the resonator, while the sidebands generated in the case of soft excitation are too small to form a pulse. We also find that hard excitation is observed when the frequency of the pumping light differs significantly from the frequency of the pumped mode, while soft excitation is observed for the nearly-resonant pumping. We compare  results of our numerical simulations with the predictions of the analytical model developed for the comb description \cite{matsko11ol}, and find that the simulation predicts formation of wider frequency combs and shorter optical pulses relative to the analytically found values. The numerical simulations also show that the comb spectral width grows slower that the linearized simulations predict \cite{chembo10ol}.

Let us consider the FWM process that involves only three resonant modes. The  evolution of the mode amplitudes is described by equations \cite{matsko05pra}
\begin{eqnarray} \nonumber
\dot {A}+\Gamma_0 A = && ig \bigl [ |A|^2 + 2 |B_+|^2 + 2 |B_-|^2 \bigr ] A+  \\  &&  2igA^* B_+B_-  + F_0, \label{Aeq}
\\  \nonumber
\dot {B}_+ + \Gamma_{+}B_+ = && ig \bigl [2|A|^2 + |B_+|^2 + 2 |B_-|^2 \bigr ] B_+ + \\  &&
  igB_-^* A^2, \label{Bpeq}
\\  \nonumber
\dot {B}_- + \Gamma_- B_-  =&&  ig \bigl [2|A|^2 + 2|B_+|^2 + |B_-|^2  \bigr ] B_-  + \\  &&
igB_+^* A^2, \label{Bmeq}
\end{eqnarray}
where $\Gamma_{0} =  i (\omega_0 -\omega) + \gamma_0$ and $\Gamma_{\pm} =  i (\omega_\pm - \tilde \omega_\pm) + \gamma_\pm$, $g=\hbar \omega_0^2 c n_2/({\cal V}n_0^2)$ is the coupling parameter \cite{matsko05pra}, ${\cal V}$ is the mode volume, $n_2$ is the nonlinearity, $F_0=(2 \gamma_0 P/(\hbar \omega_0))^{1/2}$ describes the amplitude of the continuous wave external pump, $P$ is the pump power;  $A$, $B_+$, and $B_-$ are the slow amplitudes of the pump and sidebands respectively.  Decay rates $\gamma_0$, $\gamma_+$, and $\gamma_-$ reflect both coupling and intrinsic losses of the modes. We assume that the modes are overloaded, i.e. the loss primarily results from the coupling.
\begin{figure}[htbp]
  \centering
  \includegraphics[width=7.cm]{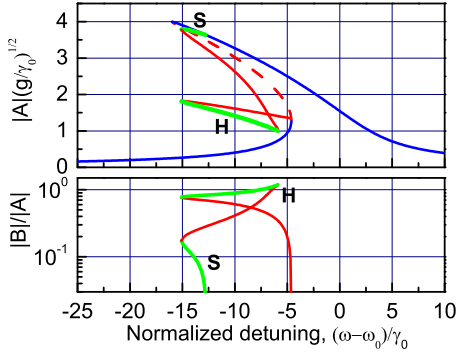}
\caption{ Normalized amplitude of the light within the optically pumped mode, and the relative amplitude of the oscillation sidebands vs pumping frequency. Stability regions corresponding to  soft ({\bf S}) and hard ({\bf H}) excitation are shown by green solid lines. Unstable solutions are depicted by red solid lines. The amplitude of the field within the optically pumped mode with no sidebands generated ($B \equiv 0$) is shown by the blue solid line and red dashed line. The dashed line stands for the unstable solution for the amplitude of the pumped mode when the modulation sidebands are absent.} \label{fig1}
\end{figure}

To keep a connection between our mathematical models and the real physical world we consider  oscillations in a magnesium fluoride whispering gallery mode resonator of 340~$\mu$m radius (approximately 100~GHz FSR), similar to the resonator studied in \cite{wang11arch}. The fundamental TE mode of the resonator is pumped with 1721~nm light. The modes of the resonator have $200$~kHz loaded full width at half maximum, and the GVD of the modes results in the condition $2\omega_0-  \omega_{+}-  \omega_{-} = -\gamma_0$. The cubic nonlinearity is $n_2=10^{-16}$~cm$^2/$W, and the refractive index is $n_0=1.38$. The mode volume is ${\cal V} =1.3\times10^{-7}$~cm$^3$ which corresponds to the coupling constant $g=1.47\times10^{-3}$~s$^{-1}$. By selecting a pump power $P=0.235$~mW, we obtain $(F_0/\gamma_0)(g/\gamma_0)^{1/2}=4$ for the normalized pumping constant. It is worth noting that the selected value of the pump power is 16 times larger than the threshold value needed for the hyper-geometric oscillation to start.

We  solve the set (\ref{Aeq}-\ref{Bmeq}) in steady state, consider a symmetric case, i.e. put $\gamma_+ = \gamma_- = \gamma_0$, so that $|  B_+  |=|  B_-  |=|  B |$, and present the slow amplitudes of the fields as $A=| A  | (1 + \delta {\cal A})\exp[i (\phi_0 + \delta
\phi_0)]$ and $B_\pm=| A  |( {\cal B}   + \delta {\cal B}_\pm)
\exp[i (\phi_\pm + \delta \phi_\pm)]$, where $\delta {\cal A}$, $\delta {\cal B}_+$, and $\delta {\cal B}_-$ stand for the amplitude deviations of the fields related to the drive amplitude, and $\delta \phi_0$, $\delta \phi_+$, and $\delta \phi_-$ stand for the phase deviations of the fields. We substitute these expressions into (\ref{Aeq}-\ref{Bmeq}), linearize the set of equations in the vicinity of the steady state solution, and study its eigenvalues. The steady state solution is considered as stable if all the eigenvalues are negative.
\begin{figure}[htbp]
  \centering
  \includegraphics[width=8.5cm]{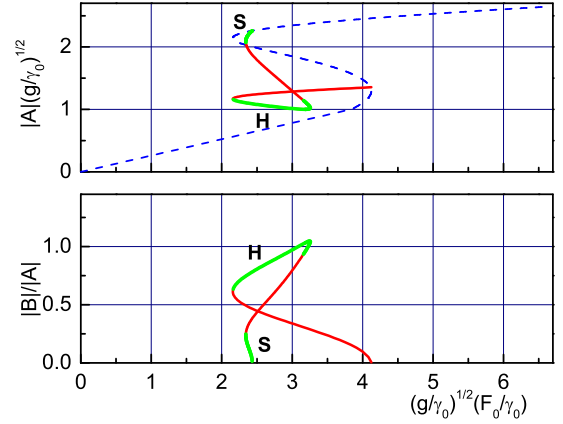}
\caption{ Normalized amplitude of the light within the optically pumped mode and the relative amplitude of the oscillation sidebands vs normalized amplitude of the pumping light. The detuning of the pumping light from the corresponding optical mode is fixed at $\omega=\omega_0-4.7 \gamma_0$ (the mode is pumped at its red wind). Dashed blue line stands for the undisturbed solution for the light accumulated in the pumped mode, $B \equiv 0$. Red and green solid lines describe unstable and stable steady state solutions respectively.} \label{fig2}
\end{figure}

The results of calculations are shown in Figs.~(\ref{fig1}) and (\ref{fig2}). We have found that there are two regions where the stable hyper-parametric oscillation exists. One solution occurs in the vicinity of the peak of the resonant curve, shifted due to  self-phase modulation effect. The oscillation belonging to this branch can be excited adiabatically if one slowly reduces the frequency of the pump laser approaching the optical resonance. It also can be excited if one fixes the laser frequency at the red wing of the resonance and then reduces the power of the pump (\ref{fig2}). The oscillation sidebands are always much smaller than the pump amplitude for this solution. The numerical solution of (\ref{Aeq}-\ref{Bmeq}) shows that the oscillation is excited when all the initial conditions are zero. Therefore, we conclude that the stability branch corresponds to the case of  soft excitation.

The other stable solution exists when the oscillation sidebands have fixed nonzero power. The stability region is localized in the space of parameters and does not cross the region of stable solution for $B \equiv 0$. Only non-adiabatic change of the parameters of the system allow reaching the localized stability region. A solution of the initial value problem (\ref{Aeq}-\ref{Bmeq}) in the approximation of zero initial conditions does not reveal the stability region. Hence, this stable branch describes the oscillation with hard excitation. The localized stable attractor, corresponding to the hard excitation regime cannot be discovered if the original set of equations describing the oscillation is solved in the approximation of small sidebands.
\begin{figure}[htbp]
  \centering
  \includegraphics[width=8.cm]{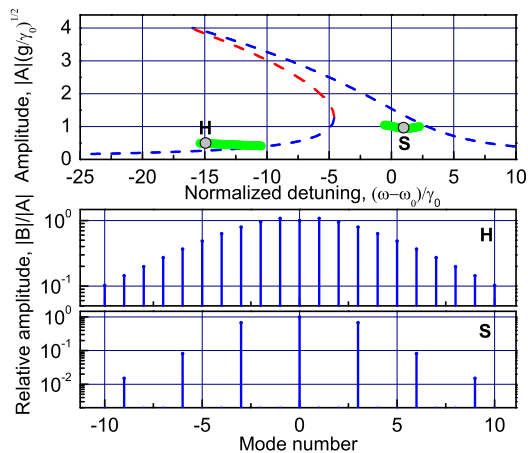}
\caption{ Normalized amplitude of the optically pumped mode and the relative amplitude of the oscillation sidebands vs pumping frequency found for the same conditions as used in Fig.~\ref{fig1}. The shown frequency combs are generated when the parameters of the system corresponding to points {\bf H} and {\bf S} are selected.} \label{fig3}
\end{figure}

The analysis that involves only two generated optical sidebands is not entirely valid for the description of  hyper-parametric oscillations in realistic resonators, since a nonlinear interaction of multiple modes should be taken into account. Only the soft-excitation regime that produces small oscillation sidebands can be simulated using a limited number of interacting modes, as the higher order oscillation sidebands are expected to have even smaller power. The hard-excitation regime, in which the oscillation sidebands can be more powerful than the light confined in the pumping mode, cannot be strictly understood from the three-mode model.

To handle this problem we have analyzed the multimode regime numerically, deriving a set of equations, similar to (\ref{Aeq}-\ref{Bmeq}), for many interacting modes, and solving this set in the steady state. With our computing capability generation of  optical frequency combs with mutually interacting 21 optical modes can be described. To take the GVD into account we assumed that for any pair of sideband modes symmetric with respect to the optically pumped mode ($\omega_0- \omega_{-} \approx \omega_{+}-\omega_0$) the mode unequidistance is defined by $2\omega_0- \omega_{+}- \omega_{-} = c\beta_2(\omega_+-\omega_-)^2/4n_0$.
\begin{figure}[htbp]
  \centering
  \includegraphics[width=7.5cm]{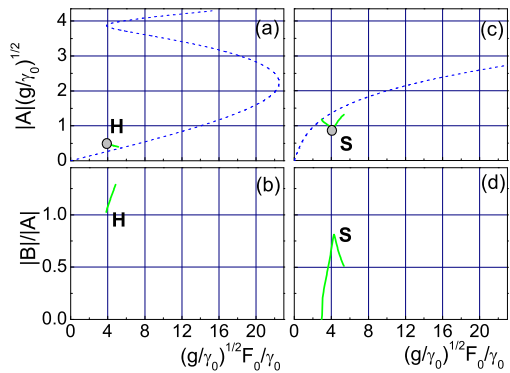}
\caption{ Normalized amplitude of the light within the optically pumped mode and the relative amplitude of the oscillation sidebands vs normalized amplitude of the pumping light. The detuning is $\omega=\omega_0-15 \gamma_0$ for plots (a) and (b) (hard excitation), and $\omega=\omega_0+0.92 \gamma_0$ for plots (c) and (d) (soft excitation). The  power for the first sideband is taken from the carrier.} \label{fig4}
\end{figure}

The numerical solution revealed excitation regimes similar to those found for the case of two optical sidebands (see Figs.~\ref{fig3} and \ref{fig4}). A broad frequency comb is generated in the occasion of the hard excitation regime. The comb harmonics have spacings equal to a single FSR of the resonator. The frequency comb resulting from  soft excitation has harmonics separated by three FSRs. There is no single-FSR comb characterized with soft excitation for the selected pump power. It is possible to generate the frequency comb with harmonics separated by double or single FSR if smaller optical power is selected.
\begin{figure}[htbp]
  \centering
  \includegraphics[width=8.5cm]{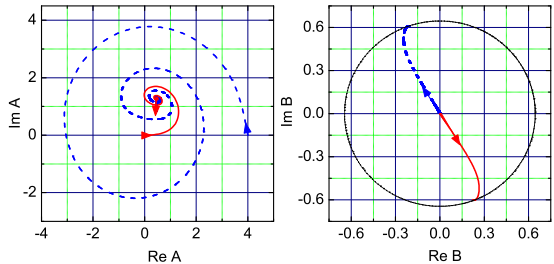}
\caption{ Phase diagram for the pump and the first  oscillation sideband in the case of soft excitation. The blue dashed line corresponds to nonzero initial conditions, and the red solid line to zero initial conditions. The system converges to the same steady state solution with different phase of the oscillation sideband. } \label{fig5}
\end{figure}

\begin{figure}[htbp]
  \centering
  \includegraphics[width=8.5cm]{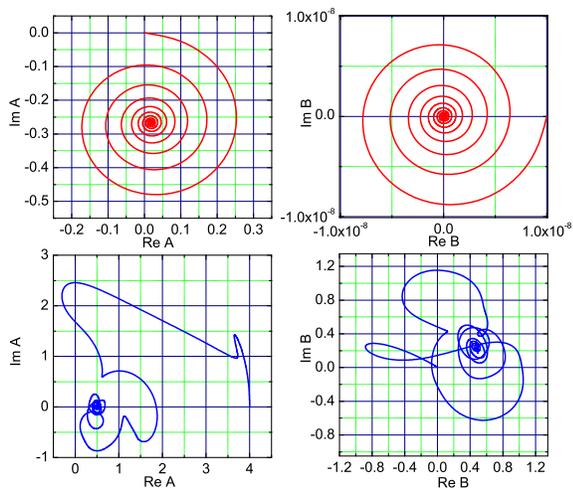}
\caption{ Phase diagram for the pump and the first  oscillation sideband in the case of hard excitation. The system does not oscillate if the initial field in the pumped mode is zero (red line). The oscillation is excited if the initial value of the pump is not zero.} \label{fig6}
\end{figure}

To demonstrate the hard and soft excitation regimes of the oscillation we studied the dynamical behavior of the light in the optical modes for zero and nonzero initial conditions (see Figs.~\ref{fig5} and \ref{fig6}). The outcome of the calculation (the steady state solution) is the same for the case of soft excitation (Fig.~\ref{fig5}). The oscillation characterized with hard excitation regime occurs only for nonzero initial conditions (Fig.~\ref{fig6}).

\begin{figure}[htbp]
  \centering
  \includegraphics[width=8.5cm]{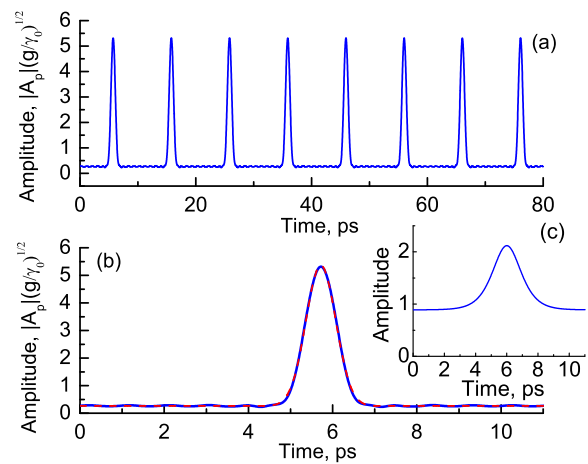}
\caption{Temporal behavior of the normalized optical field within the resonator. Sub-picosecond optical pulses are formed. Inset: the optical pulse envelope found from the analytical expression presented in \cite{matsko11ol}.  } \label{fig7}
\end{figure}

Since we can find the phase and the amplitude of the oscillation sidebands, we are able to calculate the temporal behavior of the field within the resonator. We do this for a broad frequency comb and compare the result with  predictions of the analytical method described in \cite{matsko11ol} (Fig.~\ref{fig7}). As expected, the generated harmonics create  optical pulses traveling within the resonator. The numerical modeling and the analytical theory predict slightly different behavior for the system. The reason is that the analytical model does not take into account the finite length of the path of the pulse in the resonator and requires that the pulse should be generated at the specific detuning $\omega=\omega_0-5.3 \gamma_0$. The numerical solution is unstable at this point, and instead is stable at a neighboring region of detuning values (Fig.~\ref{fig3}).

To conclude, we have developed an analytical model and numerical simulations to study the influence of  parameters of the pump light on the dynamics of Kerr comb generation. We found that  hyper-parametric oscillation and optical frequency comb generation in optical nonlinear resonators can have both hard and soft excitation regimes. We present several examples of such  behavior and show that the hard excitation regime leads to formation of short optical pulses in the resonator.

This work was supported in part by DARPA MTO (IMPACT program).



\end{document}